\documentclass[useAMS,usenatbib]{mn2e}
\usepackage{url}
\usepackage{graphicx}
\usepackage{verbatim}
\usepackage[usenames]{color}
\usepackage[flushleft]{threeparttable}
\DeclareGraphicsExtensions{.pdf,.png,.jpg,.mps,.eps,.ps}
\usepackage{amsmath}
\usepackage{natbib}
\usepackage{color}
\usepackage{graphicx}
\usepackage{lscape}
\usepackage{gensymb}

\newcommand\nustar{{\it NuSTAR}}
\newcommand\swift{{\it SWIFT}}
\newcommand\asca{{\it ASCA}}

\newcommand\sax{{\it BeppoSAX}}
\newcommand\chandra{{\it Chandra}}
\newcommand\rosat{{\it ROSAT}}
\newcommand\suzaku{{\it Suzaku}}
\newcommand\rxte{{\it RXTE}}
\newcommand\xmm{{\it XMM-Newton}}
\newcommand\inte{{\it INTEGRAL}}
\newcommand\einstein{{\it Einstein}}

\newcommand\ks{{\rm~ks}}

\newcommand\kev{{\rm~keV}}
\newcommand\ev{{\rm~eV}}
\newcommand\kms{\ifmmode {\rm~km\ s}^{-1} \else ~km s$^{-1}$\fi}
\newcommand\Hunit{\ifmmode {\rm~km\ s}^{-1}\ {\rm Mpc}^{-1}
        \else ~km s$^{-1}$ Mpc$^{-1}$\fi}
\newcommand\ctssec{\ifmmode {\rm~count\ s}^{-1} \else ~count s$^{-1}$\fi}
\newcommand\ergsec{\ifmmode {\rm~erg\ s}^{-1} \else
        ~erg s$^{-1}$\fi}
\newcommand\funit{\ifmmode {\rm~erg\ s}^{-1}\;{\rm cm}^{-2} \else
        ~ergs s$^{-1}$ cm$^{-2}$\fi}
\newcommand\phflux{\ifmmode {\rm~photon\ s}^{-1}\;{\rm cm}^{-2}
        \else   ~photon s$^{-1}$ cm$^{-2}$\fi}
\newcommand\efluxA{\ifmmode {\rm~erg\ s}^{-1}\;{\rm cm}^{-2}\;{\rm
        \AA}^{-1} \else ~erg s$^{-1}$ cm$^{-2}$ \AA$^{-1}$\fi}
\newcommand\efluxHz{\ifmmode {\rm~erg\ s}^{-1}\;{\rm cm}^{-2}\;{\rm
        Hz}^{-1} \else ~erg s$^{-1}$ cm$^{-2}$ Hz$^{-1}$\fi}
\newcommand\cc{\ifmmode {\rm~cm}^{-3} \else cm$^{-3}$\fi}
\newcommand\FWHM{\ifmmode {\rm~FWHM} \else ${\rm~FWHM}$\fi}
\newcommand\Msun{\ifmmode M_{\odot} \else $M_{\odot}$\fi}
\newcommand\Lsun{\ifmmode L_{\odot} \else $L_{\odot}$\fi}
\newcommand\ltsim{\raisebox{-.5ex}{$\;\stackrel{<}{\sim}\;$}}

\newcommand\hbeta{\ifmmode {\rm H}\beta \else H$\beta$\fi}
\newcommand\Kalpha{\ifmmode {\rm K}\alpha \else K$\alpha$\fi}
\newcommand\nh{\ifmmode N_{\rm H} \else N$_{\rm H}$\fi}
\usepackage{graphicx}
\usepackage{url}
\usepackage{verbatim}
\usepackage{color}

\title[\nustar{} \& \swift{} spectroscopy of 4U~1728--34]{\nustar{} and \swift{} joint view of neutron star X-ray binary 4U~1728--34 : Disc reflection in the island and lower banana states}
\author[Mondal et al.]{\parbox[]{6.5in}{{Aditya S. Mondal$^{1}\thanks{E-mail: adityas.mondal@visva-bharati.ac.in}$, Mayukh Pahari$^{2}$, G. C. Dewangan$^{2}$, R. Misra$^{2}$, B. Raychaudhuri$^{1}$} \\ 
  \small
$^{1}$Department of physics, Visva-Bharati, Santiniketan, West Bengal-731235, India \\
$^{2}$Inter-University Centre for  Astronomy \& Astrophysics (IUCAA), Pune, 411007 India \\
}}

\date{\today}
\begin{document}
\maketitle
\begin{abstract}
We analyze two simultaneous \nustar{} and \swift{} data of the Atoll type neutron star (NS) X-ray binary 4U~1728--34 observed on 1 and 3 October, 2013. Based on power density spectra, hardness ratio and colour-colour diagram, we infer that the first and second observations belong to the island state and lower banana state, respectively. During island state, a low-frequency quasi-periodic oscillation (LFQPO) at $\sim4.53$ Hz is observed along with four type-I X-ray bursts ($\sim 10-15$ s duration each). The X-ray luminosity of the source during the island and lower banana states are found to be $L_{X}= 1.1$ and $1.6\times10^{37}$ erg s$^{-1}$, respectively which correspond to $\sim 6\%$ and $\sim 9\%$ of the Eddington luminosity. The combined burst spectrum is well represented by a blackbody with a characteristic temperature of $2.22\pm0.05\kev$ and the blackbody radius of $9.87\pm1.86$ km which is consistent with the typical radius of the neutron star. The persistent energy spectra from both observations in the energy band $1-79\kev$ are well described with thermal emission from the NS surface ($kT_{bb}\simeq1-2.5\kev$), Comptonized emission of thermal seed photons from the hot boundary layer/corona and the strong reflection component from the accretion disc. We detect a broad Iron line in the $5-8 \kev$ band and reflection hump in the $15-30 \kev$ band. These features are well modelled by the {\tt relxill} reflection model. From joint spectral fitting, we constrain the inclination angle of the binary system and inner disc radius to be $22\degree-40\degree$ and $(2.8-4.3)\times R_{ISCO}$, respectively. From inner disc radius, we estimate the magnetic field to be $(3.3-6.5)\times10^{8}$ Gauss.   
\end{abstract}
\section{introduction}

A neutron star (NS) low mass X-ray binary (LMXB) has a low mass companion star ($\ltsim 1 M_{\odot}$) from which it accreates matter during the course of evolution. The transfer of matter towards the compact object generally occurs via accretion disc in which the matter moves in near Keplerian orbits. Guided by the observational evidences, the Keplerian disc in many models is terminated at some larger inner disc radii ($R_{in}$) by radiation drag or strong magnetic field. Depending upon the mass accretion rate, LMXBs exhibits variation in their spectral characteristics. Sometimes the hot surface of the NSs may be directly visible in the X-ray band and this can be used as an important probe to determine the NS radii. In the last few years, asymmetric Iron K$\alpha$ lines originating from the inner part of the accretion disc has been observed from the persistent spectra of many NS LMXBs \citep{2007ApJ...664L.103B, 2008ApJ...674..415C, 2008ApJ...688.1288P, 2009A&A...493L..39P, 2009MNRAS.399L...1R}. The broad Iron line observed in NS LMXBs are generally interpreted in terms of reflection of the primary X-ray continuum on the inner accretion disc. These are one of the most powerful probes to infer the properties of the plasma in the innermost part of the accretion disc around a compact object.\\

NS LMXBs are classified into two categories: Atoll and Z-sources, depending upon the track described in the X-ray colour-colour diagram or hardness-intensity diagram \citep{1989A&A...225...79H}. In the colour-colour diagram of atoll sources, two branches are usually observed, the island and the banana (banana branch is again sub categorized as lower banana and upper banana). The island branch is associated with low mass accretion rate and the source usually show characteristics of the low/hard state. In this state, it is generally believed that the accretion disc is truncated relatively far away from the compact object. As the mass accretion rate increases, the source moves from the island to the banana branch. It is thought that the accretion disc approaches the compact object and the source spectra usually show soft state characteristics. While moving from island to banana state, the temperature of the corona generally decreases and thermal emission in the form of blackbody/disc blackbody increases \citep{1995xrbi.nasa..252V}. \\
 
\begin{figure*}
\centering
\includegraphics[scale=0.30,angle=-90]{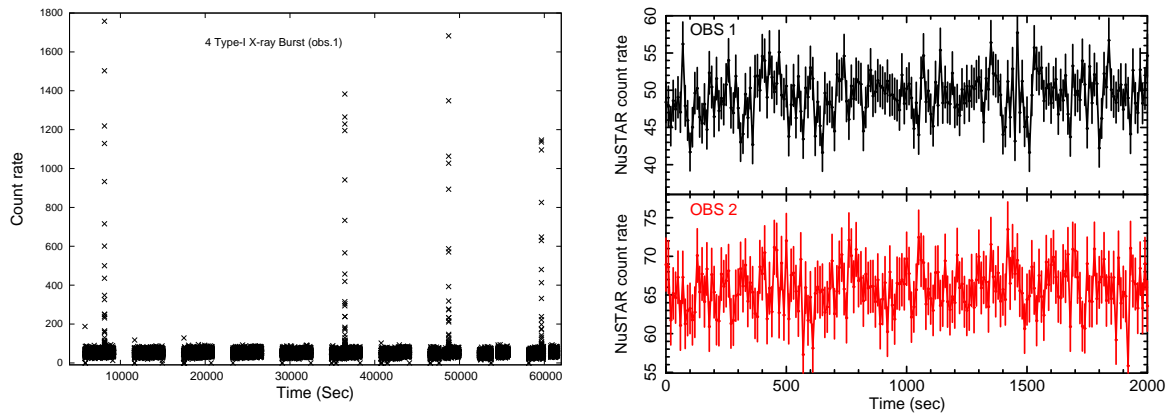}
\includegraphics[scale=0.30,angle=-90]{fig1b.ps}
\caption{The left panel shows the plot of 3-79 keV, background subtracted \nustar{} lightcurve from the OBS 1 of 4U~1728--34 with the lightcurve bin size 0.75 sec. Four type-I X-ray bursts have been detected in the light curve. The light curve of the persistent emission (for both the observation) is shown in the right panel when no bursts have been detected.}
  \label{fig_lc}
\end{figure*}

4U~1728--34 is an atoll type low mass X-ray binary and possibly an ultracompact X-ray binary with an evolved, Hydrogen poor, donor star \citep{2003ApJ...593L..35S}. The compact object of this binary is a weakly magnetized accreting neutron star and its optical counterpart is not identified yet. It was discovered in 1976 with the astronomy satellite SAS-3 \citep{1976IAUC.2922....1L, 1976ApJ...210L..13H}. Type-I X-ray burst as well as double peaked burst profile have been observed from this source \citep{1984ApJ...281..337B, 1976ApJ...210L..13H}. Type-I burst morphology indicates Helium rich bursts \citep{2008ApJS..179..360G}. Previously, several authors used the Eddington luminosity of burst profile to constrain the distance of the source in the range 4.1 to 5.1 kpc \citep{2000ApJ...542.1034D, 2003ApJ...590..999G}. The source also showed the presence of kilo-hertz quasi-periodic oscillations (kHz QPO's) in the persistent emission and a nearly coherent oscillation at $\sim 363$ Hz \citep{1996ApJ...469L...9S} which has been considered to be the spin frequency of the neutron star.    \\
 
It is well known that the spectra of the NS LMXBs can be described with the different combination of models. So far, for the soft X-ray range, the two main models widely used are `eastern model' by \citet{1984PASJ...36..741M, 1989PASJ...41...97M} and `western model' by \citet{1986MNRAS.218..129W}. The `eastern model' consists of a multi-temperature disc component and a single-temperature blackbody component originating from the NS surface or the boundary layer between the disc and the NS. On the other hand, the `western model' uses a single-temperature blackbody and a Comptonized component which is the inverse Compton scattering of the soft seed photon from the disc by the hot energetic electrons of the corona. In the soft spectral state, the soft thermal component dominates while during the hard spectral state soft component decreases and Comptonization component becomes predominant. In the past years, spectral analysis of the source has been performed using the data from the satellites like \einstein{} \citep{1981ApJ...247L..17G}, \rosat{} \citep{1999ApJ...511..304S}, \asca{} \citep{2001ApJ...547..420N}, \sax{} \citep{2000A&A...360L..35P, 2000ApJ...542.1034D}, \chandra{} \citep{2006A&A...448..817D}, \inte{} \citep{2006A&A...458...21F}, \rxte{} \citep{2006A&A...448..817D} and \xmm{} \citep{2010A&A...522A..96N, 2011A&A...530A..99E}. It is one of few NSXBs which is detected in Radio band and shows a hard powerlaw tail during its island state \citep{2003MNRAS.342L..67M, 2007ApJ...654..494T}. Two possible origins non-thermal tail has been anticipated : (1) hard tail is caused by the Comptonization of disc photons in the presence of thermal/non-thermal hybrid population of hot electron near the compact object \citep{1998PhST...77...57P}. This model successfully accounts the origin of hard tail in 4U 1820-30 \citep{2007ApJ...654..494T} (2) Hard X-ray photons in the tail may also originate from the base of the jet \citep{2005ApJ...635.1203M} or the bulk motion Comptonization model \citep{1998ApJ...493..863T}. 

Apart from the continuum processes, the hard X-ray photons can be reflected off the inner edge of the accretion disc and give rise to reflection spectra which allow for new views of the accretion geometry in LMXBs. The most important features of the disc reflection in X-ray binaries are: (1) a broad emission line in the Fe-K band ($6.4 - 6.97 \kev$) and (2) a Compton back-scattering hump peaking at $\sim 15-30\kev$ \citep{1989MNRAS.238..729F}. Broad emission lines $\sim 6-7\kev$ has been detected in the X-ray spectra of 4U~1728--34. \citet{2011A&A...530A..99E} reported the broad Iron line from the spectral analysis of a \xmm{} observation and interpreted this as the emission from the highly ionized Iron, which could come from the ionized inner accretion disc or from a strongly ionized corona. They fitted the continuum with the thermal Comptonized models, {\tt compTT} and {\tt nthcomp} separately and no blackbody component was statistically required. To fit the Iron line profile, they used different line models like {\tt diskline}, {\tt rellline} as well as the self-consistent relativistically smeared reflection model. Broadband \sax{} ($0.1-200\kev$) and \rxte{} ($3.0-60\kev$) spectra of 4U~1728--34 was well described by a two component model - a soft component produced by the disc and a Comptonized spectrum responsible for the hard emission ( \citealt{2000ApJ...542.1034D, 2000A&A...360L..35P}). \citet{2006A&A...448..817D} performed the spectral analysis of the source using a simultaneous \chandra{} and \rxte{} observation taken on March 2002. They interpreted the residual seen  within $\sim6 - 9\kev$ energy band as the presence of two absorption edges of ionized Iron at $7-9 \kev$. \citet{2006A&A...458...21F} fitted broadband spectrum of $3 - 200\kev$ obtained from the \inte{} data with a multicolor disk blackbody and a Comptonized emission.  The \inte{} spectra showed no reflection features, but a change of spectral shape between soft and hard state was reported. Therefore, 4U~1728--34 shows variable reflection which can be caused by the detection of the source in two different states. To fit the spectra obtained from \inte{} and \rxte{} data, \citet{2011MNRAS.416..873T} used several combinations of models for different spectral states. Hard state spectra showed the presence of residuals at high energy ($>50\kev$) which was modelled either with a power-law or a Comptonization component. They used a Gaussian component to fit the broad emission line after fixing the line energy at $6.5\kev$ and constraining line width at $\sim0.3-0.7\kev$.  \\       

In this work, for the first time, we present a detailed broad band study of the source 4U~1728--34 using \nustar{} and \swift{} spectra in the energy band $1-79\kev$ during the island and the lower banana branch. \nustar{} observation shows type-I X-ray bursts from the NS surface. We perform spectral analysis of these bursts and estimate the radius of the NS. We also study the broadband spectrum during the non-burst period. Here we investigate the origin and the nature of the broad Iron line observed $\sim5-8\kev$ and Compton hump in the $15-30\kev$ energy band in the light of the disc reflection spectra modelling.  
 \\

The paper is organised as follows: We describe the observations and the detail of data reduction in sec.2. In sec.3 and sec.4 we describe the temporal and spectral analysis, respectively. Finally, in sec.5 we discuss our findings. We quote the uncertainties on model parameters at the $90\%$ confidence level.  

\begin{figure*}
\centering
\includegraphics[width=5.0cm,angle=-90]{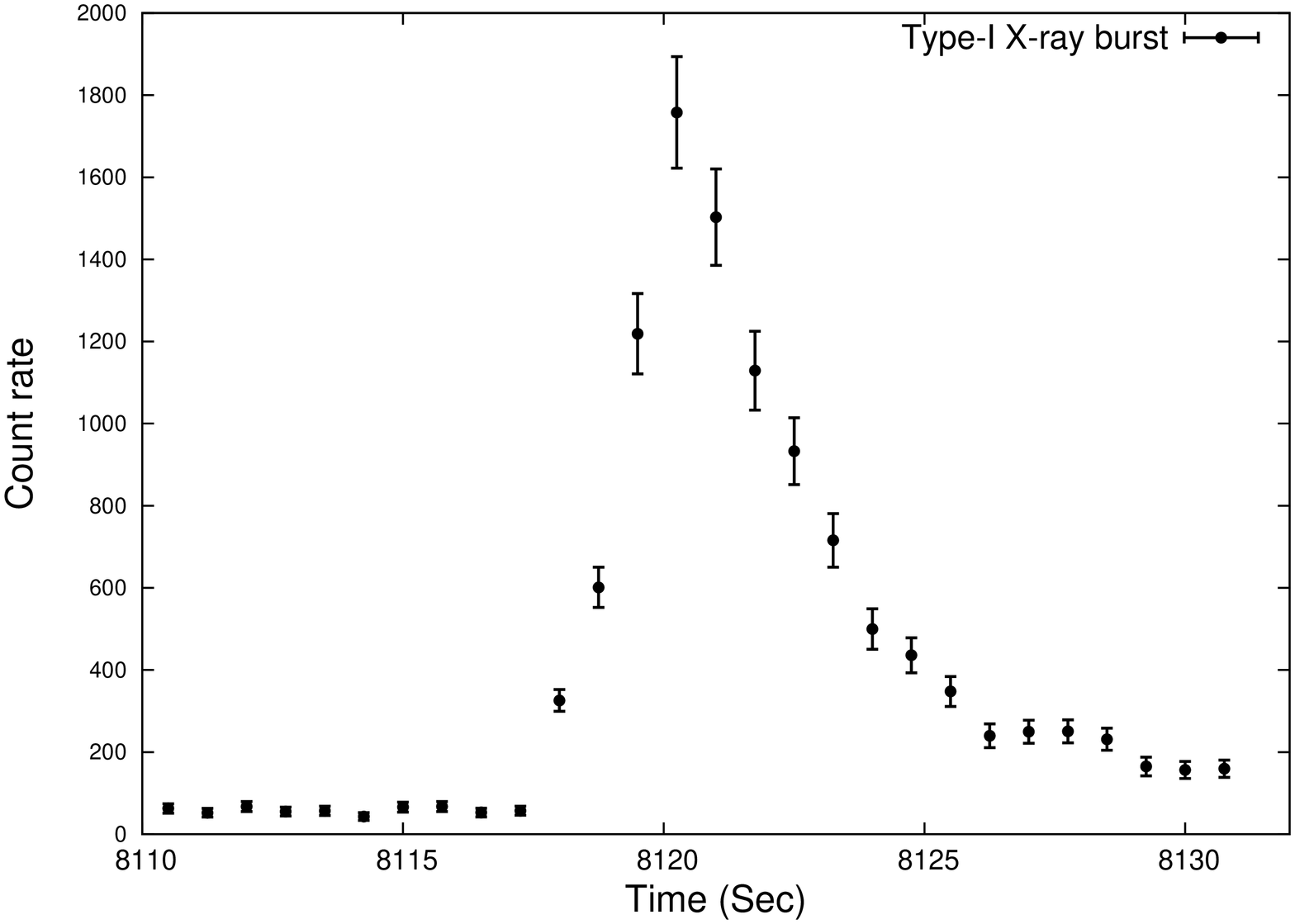}
\includegraphics[width=5.0cm,angle=-90]{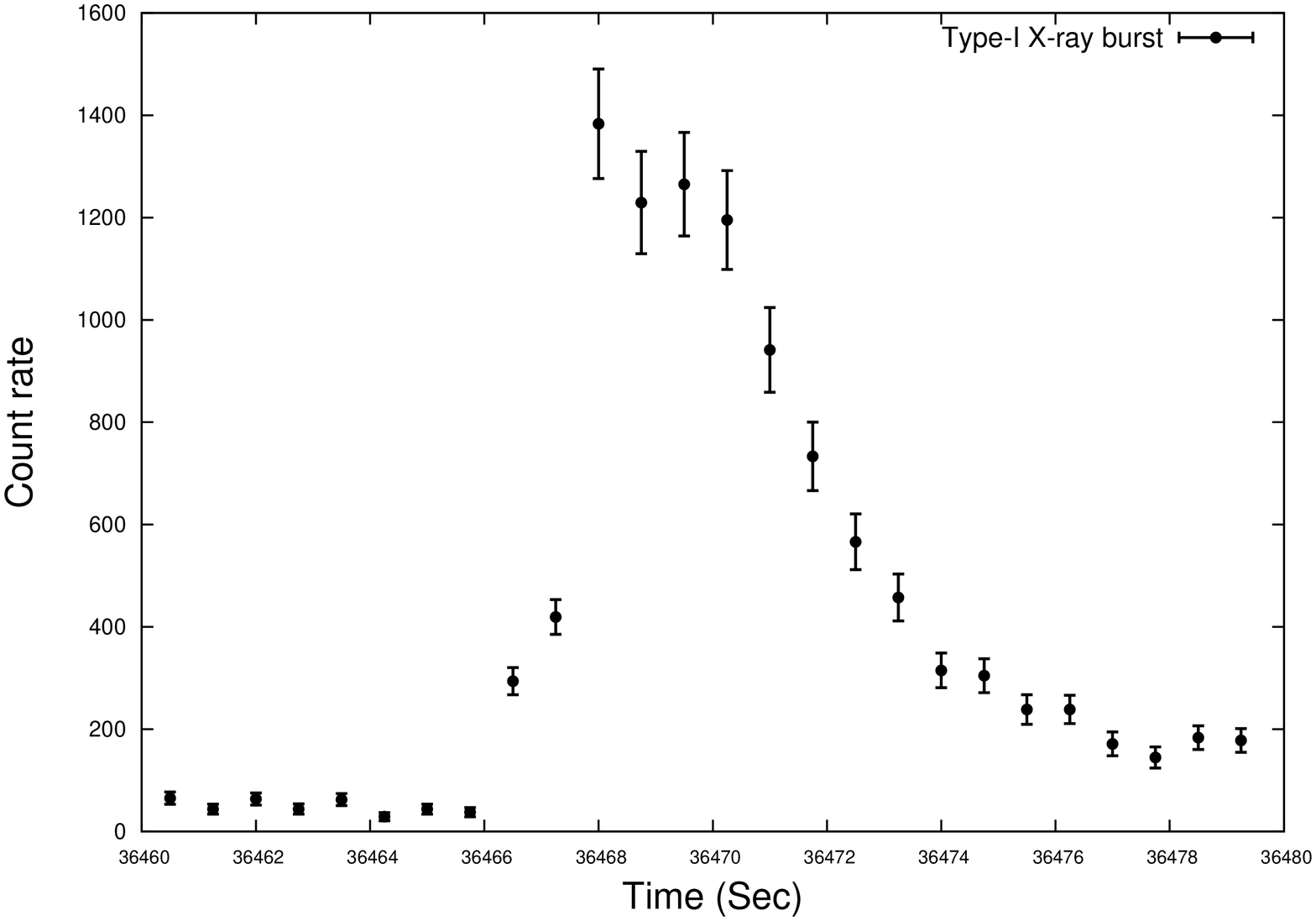}
\includegraphics[width=5.0cm,angle=-90]{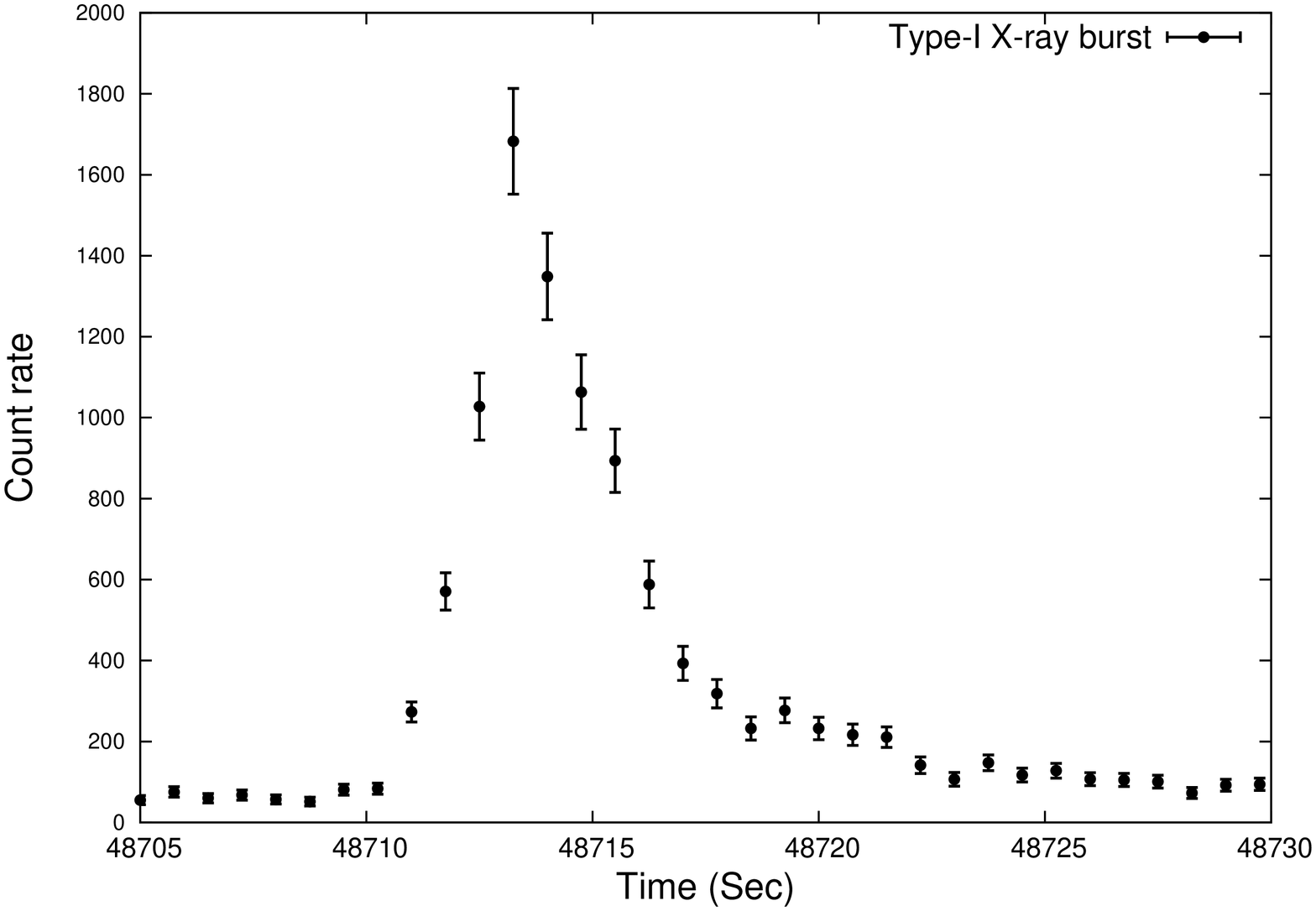}
\includegraphics[width=5.0cm,angle=-90]{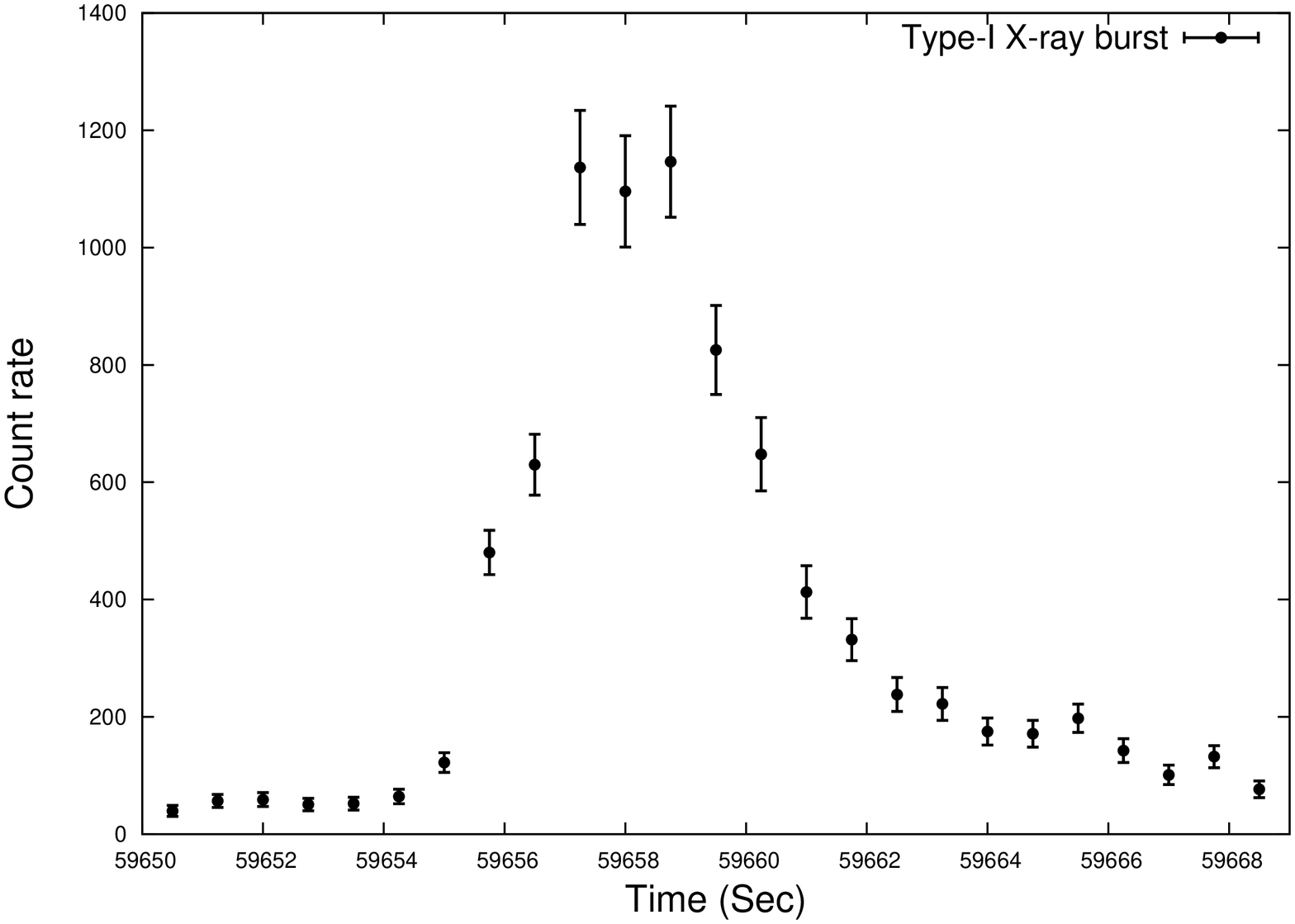}
\caption{Light curve of the individual type-I X-ray burst detected in the first \nustar{} observation of 4U~1728--34 are shown in four panels. It may be noted that the 2nd and 4th X-ray bursts have significantly broader, plateau-like peak profile with low peak counts while the 1st and 3rd bursts have very sharp peak profiles with high peak counts. }
  \label{burst_lc}
\end{figure*}    

\section{observation and data reduction}
\subsection{\nustar{} FPMA and FPMB}
\nustar{} observed the source 4U~1728--34  on 2013 October 1 and October 3 (Obs. ID: $80001012002$ and $80001012004$). NuSTAR data were collected with the focal plane module telescopes (FPMA and FPMB). The net exposure after instrumental deadtime correction are $\sim27.3$ and $\sim26.3$ ks, respectively. Table~\ref{obs_summary} lists the observation details. \\
 
We performed the data reduction using the standard \nustar{}
data analysis software ({\tt NuSTARDAS v1.5.1}) with the latest calibration (version $20150904$). We employed the {\tt nupipeline} task (version v 0.4.3) to filter the event file as well as to apply the default depth correction.
We extracted a circular region with a radius of $100\arcsec$ centered on the source position to obtain source events, whereas a region of the same dimension placed at the corner of the same detector which is away from the source, was used for the background. Source and background light curves and spectra were extracted from the FPMA and FPMB data and response files were generated employing the {\tt nuproduct} task. The data from the two FPM were modelled simultaneously in order to minimize systematic effects. Using {\tt GRPPHA}, we grouped the spectral counts using $200$ counts per spectral bin. 

\subsection{\swift{}/XRT}

The source was also observed by \swift{}/XRT simultaneously with the \nustar{} observation. 4U~1728--34 was observed for $2.1\ks$ and $1.9\ks$ on 2013 October 1 and October 3, respectively, with the XRT operated in Windowed Timing (WT) mode, the details of which are provided in Table~\ref{obs_summary}. Following standard procedure for filtering and screening, \swift{}/XRT data were reduced using the {\tt xrtpipeline v 0.13.1} tool. The background subtracted average count rate in this mode was found to be $\sim20$ counts/s, which is well below the prescribed photon pile-up limit of the WT mode data ($\sim$ 100 counts/s; \citealt{2006A&A...456..917R}). Using {\tt XSELECT v 2.4c} we extracted source and background events from a rectangular box of 40 pixels long and 15 pixels wide . The latest \swift{}/XRT
spectral redistribution matrices (RMF ver. 15) were taken from the calibration database and the {\tt xrtmkarf} tool was used
along with the exposure map file to generate an auxiliary response file (ARF) for the current observations.

\begin{table*}
\caption{Observation details for the \swift{} and \nustar{} on 4U~1728--34} 
  \begin{tabular}{|p{2.2cm}|p{2.0cm}|p{2.0cm}|p{1.5cm}|p{1.5cm}|p{1.5cm}|}
    \hline
    \hline
    Instrument & Obs. ID. & Obs. start date (dd/mm/yyyy) & Effective exposure time (ks) & Count rate (cts/s) & Obs. Mode  \\ 
    \hline
    NuSTAR/FPMA, FPMB   & 80001012002 & 01/10/2013 & 27.3  & $51\pm4$ & SCIENCE \\
    NuSTAR/FPMA, FPMB   & 80001012004 & 03/10/2013 & 26.3  & $67\pm3$ & SCIENCE \\
	SWIFT/XRT & 00080185001 & 01/10/2013 & 2.1 & $20\pm1$ & WT \\
        SWIFT/XRT & 00080185002 & 03/10/2013 & 1.9 & $16\pm1$ & WT \\
    \hline 
  \end{tabular}\label{obs_summary} \\
\end{table*}

\section{Temporal analysis}

\subsection{X-ray bursts and persistent lightcurve}
 First \nustar{} observation was performed on 1 October, 2013 with an exposure of $\sim$ 62 ks (Obs Id:80001012002, hereafter OBS 1) . The left panel of Fig.~\ref{fig_lc} shows the background subtracted lightcurve of 4U~1728--34 in the energy band $3 - 79\kev$ obtained with the \nustar{}/FPMA from the OBS 1. The source was detected with an average persistent intensity of $48\pm3$ counts $s^{-1}$. On four occasions, a rapid rise in the intensity was observed ($\sim$ by a factor of $30-35$) in the light curve of 4U~1728--34 which last for $10-15$ sec. These bursts are known as type-I X-ray bursts. These bursts were registered at $\sim8.1\ks, \sim36.4\ks, \sim48.7\ks, \sim59.6\ks$ during the OBS 1. The second observation was performed on 3 October, 2013 with an exposure of $\sim61$ ks (Obs Id: 80001012004, hereafter OBS 2). The background-subtracted, average persistent intensity of the OBS 2 was $64\pm4$ counts $s^{-1}$. Therefore, $\sim$ 33\% increase in the count rate was observed while going from the OBS 1 to the OBS 2. 2 ks lightcurve from both the observations are shown in the right panel of Fig.~\ref{fig_lc}. Only one type-I X-ray burst of $15$ sec duration was detected at $1.3\ks$ from the OBS 2. Among four type-I bursts in the OBS 1, the average peak count rate (measured by \nustar{} FPMA/FPMB) alternated between $1800$ counts s$^{-1}$ and $1300$ counts s$^{-1}$, respectively. Lightcurves of four burst profile are shown in four panels of Fig.~\ref{burst_lc}. The start and end time of the bursts were determined when the flux was $\sim 10\%$ of the peak above the persistent emission label. The X-ray bursts usually peaked $\sim 6$ sec after the initial start.  

\subsection{Power density spectra}

\begin{figure*}
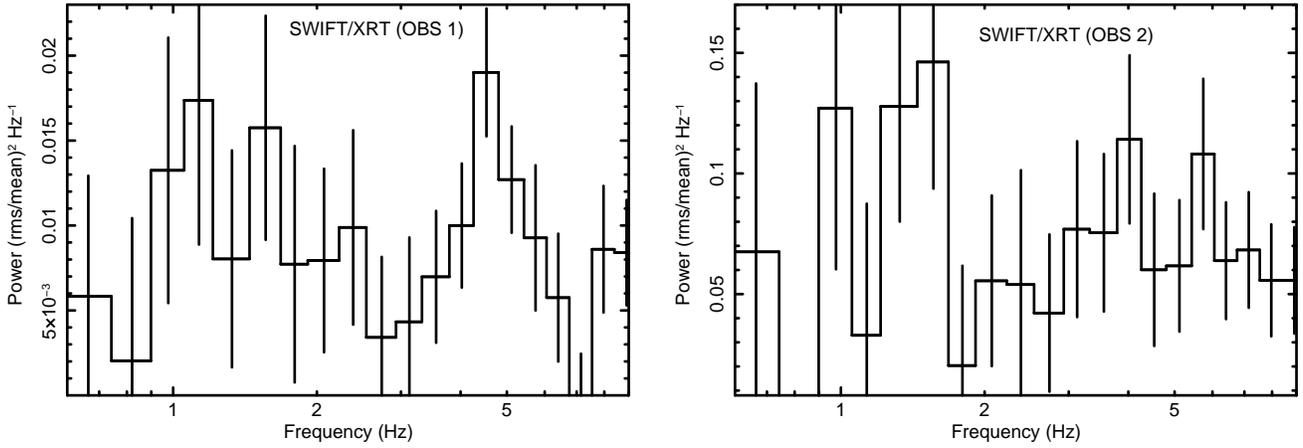

\centering
\includegraphics[scale=0.35,angle=-90]{fig3a.ps}
\includegraphics[scale=0.35,angle=-90]{fig3b.ps}
\caption{RMS normalized, Poisson noise subtracted power density spectra obtained using \swift{}/XRT data in the 0.5-10 keV energy range are shown during OBS 1 (left panel) and OBS 2 (right panel). A low-frequency QPO at $\sim$ 4.53 Hz is observed in the PDS during OBS 1 while the PDS of OBS 2 is noise dominated.}
  \label{pds}
\end{figure*}

We extracted power density spectra (PDS) from both the observations using \swift{}/XRT lightcurves in the energy band $0.5-10 \kev$. Lightcurves from both the \swift{}/XRT observations consist of two $\sim$ 1 ks continuous stretch of lightcurve with a gap between them. To avoid the gap, we treated each stretch of lightcurve separately to compute PDS. PDS were rms normalized and white noise subtracted. Both left and right panels of Fig.~\ref{pds} show PDS from the continuous part of the lightcurve from both the observations. A low-frequency quasi-periodic oscillation (LFQPO) is observed at $\sim$ 4.53 Hz in OBS 1. This is a typical characteristic of the source when it passes through the island state \citep{2001ApJ...546.1107D}. During OBS 2, no QPO is detected in the PDS. Lack of low-frequency QPO is the characteristic of banana state \citep{2001ApJ...546.1107D}.   

\subsection{Hardness ratio and colour-colour diagram}
 
\begin{figure*}
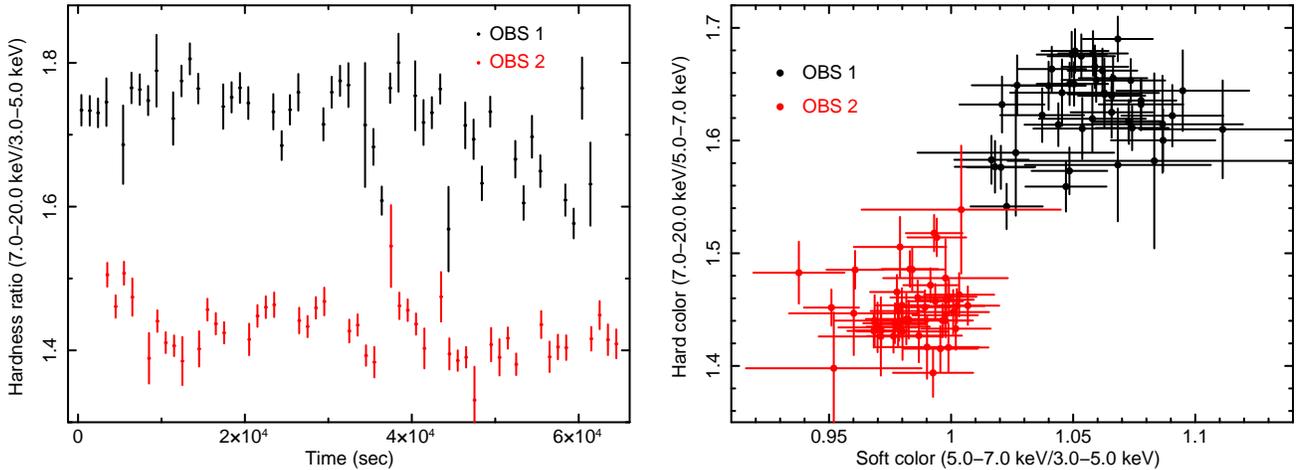

\centering
\includegraphics[scale=0.35,angle=-90]{fig4a.ps}
\includegraphics[scale=0.35,angle=-90]{fig4b.ps}
\caption{Left: Hardness ratio as a function of time is shown for OBS 1 (black) and OBS 2 (red) using \nustar{} data of 4U~1728--34. The hardness is defined as the count rate ratio of 7-20 keV and 3-5 keV. Right panel shows the colour-colour diagram during OBS 1 (black) and OBS 2 (red) using \nustar{} data. Comparing colour-colour diagram with earlier work, it is clear that OBS 1 belongs to island state while OBS 2 belongs to lower banana state.}
  \label{fig_HID}
\end{figure*}

Fig.~\ref{fig_HID} shows the hardness ratio of both the observations of the source 4U~1728--34 with \nustar{}/FPMA data. 
We defined the energy range $3-5\kev$ as the soft band (S) and the energy range $7-20\kev$ as the hard band (H). We calculated the ratio of the counts detected in the two bands (H/S) as a function of time. In the left panel of Fig.~\ref{fig_HID}, hardness ratios of the OBS 1 and the OBS 2 are denoted by black and red colours, respectively. From the hardness ratio diagram, it is clear that the \nustar{}  OBS 1 is harder compared to the OBS 2.  The hardness ratio lies in the range $1.6-1.8$ and $1.4-1.6$ for the OBS 1 and OBS 2, respectively. To ensure whether two observations belong to two different branches, we plotted colour-colour diagram (CCD) which is shown in the right panel of Fig.~\ref{fig_HID}. The soft color is defined as the ratio of count rate in $5-7\kev$ \& $3-5 \kev$ and the hard color is defined as the ratio of count rate in $7-20 \kev$ \& $5-7 \kev$. Both observations, shown by black and red circles occupy different regions in the CCD. A comparison of the shape of the CCD between our work and from \citet{2001ApJ...546.1107D}, clearly indicates that the OBS 1 belongs to the island branch while the OBS 2 belongs to the lower banana branch.

\section{spectral analysis}

\subsection{The burst spectra}
We used \nustar{}/FPMA data to produce the spectra during the bursts. We considered the beginning and the end times of each burst when the count rate was $\sim 10\%$ of the peak above the persistent emission label. For all the bursts, spectra were accumulated using central 20 sec of each burst profile as a good time interval (GTI). Initially, all the burst spectra were analyzed separately. We used a simple absorbed blackbody component ({\tt bbody} in {\tt XSPEC}) to fit  all the burst spectra and found that the best-fit parameters were similar within errors (see Table~\ref{burst_spectra}). So, in order to improve the statistics, all the burst spectra obtained from \nustar{}/FPMA (OBS 1) were added. We also extracted the persistent emission spectrum excluding the gap and outburst intervals. We subtracted the persistent emission spectrum from the combined burst spectrum by using the persistent emission spectrum as the background spectra in {\tt XSPEC}. A Similar technique was used by \citet{2000ApJ...542.1034D}, \citet{2006A&A...458...21F} to analyse X-ray burst spectra from the source 4U 1820--30 and also by \citet{2013MNRAS.436.2334P} to compute spectra from excess flux in GRS 1915+105. Thus, the resultant spectra are consisted of counts contributed solely by the thermonuclear X-ray bursts. A Similar exercise has been performed for \nustar{}/FPMB burst spectra as well.

To model individual as well as combined burst spectrum, we used a simple blackbody model ({\tt bbody} in {\tt XSPEC}) and to account for interstellar absorption we employed {\tt tbabs} model with the abundances set to {\tt wilm} \citep{2000ApJ...542..914W}, and the cross section to {\tt vern} \citep{1996ApJ...465..487V}. Due to the lack of calibrated spectra below $3 \kev$ in \nustar{}, the equivalent hydrogen column density ($N_{H}$) was not well constrained by the \nustar{} data. Therefore, we fixed the interstellar column density to $N_{H}=2.5\times10^{22}cm^{-2}$ (\citealt{2006A&A...448..817D, 2011A&A...530A..99E}). From the burst spectrum, The best-fit blackbody temperature and the radius of blackbody emission were found to be $kT_{bb}=2.2\pm0.05\kev$ and $9.87\pm1.86$ km respectively assuming the distance of 5 kpc and color correction factor of 1.7. In Table~\ref{burst_spectra} we reported the best-fit parameter values obtained by fitting the individual burst spectrum as well as the combined spectrum with the absorbed {\tt bbody} model. Combined burst spectrum along with the residuals (in units of $\sigma$) with respect to the best-fit model is shown in the left panel of Fig.~\ref{residuals}. 

\subsection{The Persistent spectra}

\subsubsection{The broadband continuum}
We simultaneously fitted spectra of \nustar{}/FPMA and FPMB in the $3.5-79\kev$ energy band and \swift{}/XRT spectra in the $1-10\kev$ energy band using {\tt XSPEC v 12.8.2} \citep{1996ASPC..101...17A}. We introduced a cross normalization factor to take into account the cross calibrations of the different instruments used. It was fixed to 1 for \nustar{}/FPMA and kept free for others instruments. It is well known that different combinations of continuum models can fit the data of neutron star LMXBs equally well \citep{2001AdSpR..28..307B}. We first attempted to fit the continuum with the typical model used for NS LMXBs of the atoll class. This model consisted of a  single temperature blackbody component ({\tt bbody} in {\tt XSPEC}) and a thermal Comptonization component {\tt compTT} \citep{1994ApJ...434..570T}, modified by the interstellar absorption modelled by {\tt tbabs} with {\tt vern} cross section \citep{1996ApJ...465..487V} and {\tt wilm} abundances \citep{2000ApJ...542..914W}. The same combination of the continuum model was previously used many times for the source 4U~1728--34 ( see e.g., \citealt{2006A&A...448..817D, 2000ApJ...542.1034D, 2006A&A...458...21F}). This continuum model, {\tt tbabs$\times$(bbody+CompTT)}, resulted in $\chi^{2}/dof$=$1693/949$ and $1992/843$ for the OBS 1 and  OBS 2, respectively (where $dof$ is the degrees of freedom). The continuum fitting yielded the best-fit electron temperature of the Comptonizing plasma of $kT_{e}\sim6-11\kev$, seed photon temperature of $kT_{seed}\sim0.5-1.1\kev$ and optical depth of $\tau\sim2-4$. To test, whether another thermal component is required or not, we added a disk blackbody ({\tt diskbb} in {\tt XSPEC}) to the existing continuum model. The addition of the {\tt diskbb} model was found to be statistically insignificant. 

However, the combination of one thermal component ({\tt bbody}) and one Compotonized component ({\tt CompTT}) provided a formally unacceptable fit, because of the presence of evident residuals at $5-8\kev$, $15-20\kev$ and small residuals at $\sim1.8\kev$ and $\sim2.3\kev$. Residuals obtained from fitting both observations are shown in the right panel of Fig.~\ref{residuals}. The most prominent reflection feature is a clear, broad Iron k$\alpha$ line profile at energies $\sim 6.4 \kev$. The observed broad excess flux in the $10-20\kev$ energy band is consistent with a Compton back-scattering hump. Previously, Compton hump peaking at $10-20\kev$ energy range has also been observed from the \nustar{} spectra of another NS LMXB Ser X-1 by \citet{2013ApJ...779L...2M}.

Before modelling reflection features, we tested other combinations of thermal and Comptonized component to fit the continuum. A combination of a cutoff power-law ({\tt cutoffpl} model in {\tt XSPEC}) and blackbody component ({\tt bbody}) provided  $\chi^{2}/dof=1824/950$ and $1927/844$ for the OBS 1 and OBS 2, respectively, with the cutoff power-law index $\Gamma=1.34\pm0.03$ and high energy cutoff value $19.1_{-0.51}^{+0.52}\kev$ (for OBS 1). 
The combination of {\tt bbody} model with another Comptonization model {\tt nthcomp} in {\tt XSPEC} provided $\chi^{2}/dof$=$1953/949$ and $1604/843$ for the OBS 1 and OBS 2, respectively with the asymptotic power-law photon index $\Gamma\sim2-2.5$ and electron temperature $kT_{e}\sim9-11\kev$ and seed photon temperature $kT_{seed}\sim1.0-1.5\kev$. With all the choice of continuum models, prominent residuals in the $\sim 5 - 8\kev$ and $\sim15-30 \kev$ energy bands were observed. Recovery of the reflection features like broad Fe K$\alpha$ emission line and Compton hump with all the continuum models indicates the significance of its presence in all measurements. \\

\subsubsection{The broad Iron line \& reflection hump}

We tried to fit the large excess seen $\sim 5 - 8\kev$ with a Gaussian line ({\tt Gaussian} in {\tt XSPEC}). We added a Gaussian model with the continuum model {\tt tbabs$\times$(bbody+CompTT)}. The addition of this model to the data improved the fit significantly for 3 additional parameters ($\Delta\chi^{2}=-485$ and $-882$ for the OBS 1 and OBS 2, respectively) and corresponding $\chi^{2}/dof$ values are $1208/946$ and $1110/841$) with line energy centre at $6.50\pm0.04$, line width $\sigma=0.78\pm0.10\kev$  and equivalent width $EW\sim135\ev$. We then tried to substitute the {\tt Gaussian} at $6.6\kev$ with a laor line profile ({\tt laor} in {\tt XSPEC}, \citealt{1991ApJ...376...90L}). This model {\tt TBabs$\times$(bbody+CompTT+laor)} did not result in considerable improvement of the fit, since it provided $\chi^{2}/dof=1231/945$ and $1224/839$ for the OBS 1 and  OBS 2, respectively. During the fitting with {\tt laor} line profile, we fixed the outer radius of the disc and the inclination angle to the value $400R_{G}$ (where $R_{G}=GM/c^{2}$ is the gravitational radius) and $60\degree$, respectively (as the inclination was not well constraint and there is no prior knowledge of source inclination, see \citealt {2011A&A...530A..99E, 2006A&A...458...21F}). This model provided an estimate of the inner radius of the disc $R_{in} \le 8 R_{G}$, line energy $E_{LAOR}=6.46^{+0.06}_{-0.03}\kev$ and emissivity index $\beta=1.8\pm0.16$. For both the observations, these parameter values were consistent within errors. The primary reason of faliure of both models to provide a satisfactory fit is the presence of Compton hump in the energy band $15-30\kev$ which cannot be modelled by the {\tt Gaussian} or by the {\tt laor} model.  \\

\begin{figure*}
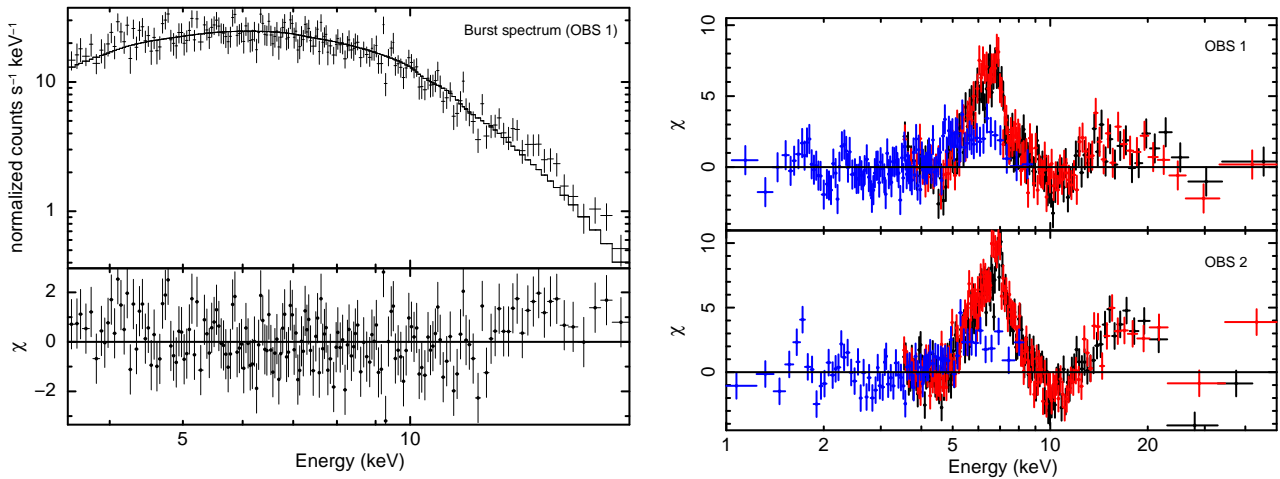

\centering
\includegraphics[scale=0.35,angle=-90]{fig5a1.ps}
\includegraphics[scale=0.35,angle=-90]{fig5b.ps}
\caption{The top left panel shows absorbed blackbody model fitted burst spectrum in the $3-20\kev$ energy range where spectra during all four type-I X-ray bursts are combined and bottom left panel shows the residuals in units of $\sigma$. Right panel shows the residual of jointly fitted \nustar{} FPMA/FPMB (red and black) and \swift{}/XRT (blue) spectra using a simple phenomenological model {\tt tbabs$\times$(bbody+CompTT)} for both the observations in the energy range $1-79\kev$. Residuals are shown in units of $\sigma$. A broad emission line feature in the energy range $5-8\kev$ is strongly detected. Above $10\kev$ a broad, hump-like excess is also visible. For better visibility, additional spectral binning by a factor of 80 is used than the original bin used for spectral fitting. }
\label{residuals}
\end{figure*}

The broadening of the line seen $\sim6.5\kev$ and the presence of other emission features (see Fig.~\ref{residuals}, Right panel) are highly suggestive for fitting the broad-band spectrum with a self-consistent reflection model.
Therefore, to fit the spectra, we applied the reflection model {\tt relxill} \citep{2014ApJ...782...76G}, which calculates disc reflection features due to an irradiating power-law source. It combines the relativistic convolution kernel {\tt relconv} \citep{2010MNRAS.409.1534D} with the reflection grid {\tt xillver} \citep{2013ApJ...768..146G}, in which reflection spectrum is chosen for each relativistically calculated emission angle rather than averaged. The fit parameter of the {\tt relxill} model are the two emissivity indices, the breaking radius $R_{br}$ where the emissivity changes, the dimensionless spin $a$, the binary inclination $i$, the inner and outer disc radii $R_{in}$ and $R_{out}$, the ionization parameter log$\xi$, the Iron abundance $A_{Fe}$, the reflection fraction $R_{refl}$, the normalization $N$ and the index $\Gamma$ and high-energy cut-off $E_{cut}$ of the power-law. \\

We first employed two component spectral model using blackbody emission from the NS surface along with the self-consistent {\tt relxill} model. This model provided fit to the joint spectra from OBS 1 and OBS 2 with $\chi^2$/dof =$1140/944$ and $1088/839$ respectively. However, many spectral parameters from both OBS 1 and OBS 2 can not be constraint and significant residual around $3-6 \kev$ exists in the fitted spectra even when reflection features are properly taken care of. This necessitates the addition of another component with the previous modelling. We introduced multi-temperature disk blackbody emission ({\tt diskbb} in {\tt XSPEC}) along with the combination of single temperature blackbody from the NS surface and {\tt relxill} reflection model. However, this did not improve the fitting and reduced $\chi^2$ remains similar.
This implies that the shape of the real incident spectra that cause the reflection from the disc is very different than an incident power-law.  Therefore, we introduced a thermal Comptonization model {\tt CompTT} in {\tt XSPEC} as a third component to the combination of blackbody and reflection models. We tied the cutoff power-law energy of the incident spectra in {\tt relxill} to that of the Comptonizing electron temperature ($\sim3 kT_{e}$). However, it may be noted that the reflection amplitude ($R_{refl}$) predicted by the best-fit {\tt relxill} model may not be the correct one due to the introduction of a thermal Comptonization model. \\

The use of thermal blackbody and thermal Comptonization models along with the self-consistent reflection models like {\tt reflionx} \citep{2005MNRAS.358..211R}, {\tt xillver} \citep{2013ApJ...768..146G} and {\tt relxill} is not a new approach. Previously, during the modelling of broadband spectra from both neutron star as well as black hole X-ray binaries, the combination of thermal emission from NS surface/BH accretion disk, thermal Comptonization and self-consistent reflection models have been used. For example, \citet{2015MNRAS.449.2794D} showed that the broadband \suzaku{} spectra in the energy range $0.7-200 \kev$ from the Atoll-type NS LMXB 4U~1705--44 (the characteristics of which is similar the source analyzed in this work) can be best described by thermal blackbody {\tt bbody in XSPEC}, thermal Comptonization {\tt nthcomp in XSPEC} and self-consistent reflection model like {\tt reflionx} or {\tt relxill}. In another approach, while modelling the hard state spectra from the BHXB GX~339--4 observed with \xmm{}, \citet{2016MNRAS.458.2199B} found that along with the disk blackbody {\tt diskbb in XSPEC} and the currently best Compton reflection model {\tt relxill}, a Comptonization model {\tt nthcomp} is also required to fit the spectra in the energy range $0.3-10 \kev$. They found that setting the disk truncation radius to that of the reflector leads to unphysical behaviour of inner disk radius. Additionally, they discussed inaccuracies introduced to the reflection fraction due to normalization of the  incident and reflected flux in 20-40 keV energy range. We found that the combination of thermal blackbody, thermal Comptonization and self-consistent reflection model fitted our spectra during island and lower banana branch observations very well. Spectral parameters are consistent with earlier estimation. \\

\begin{figure*}
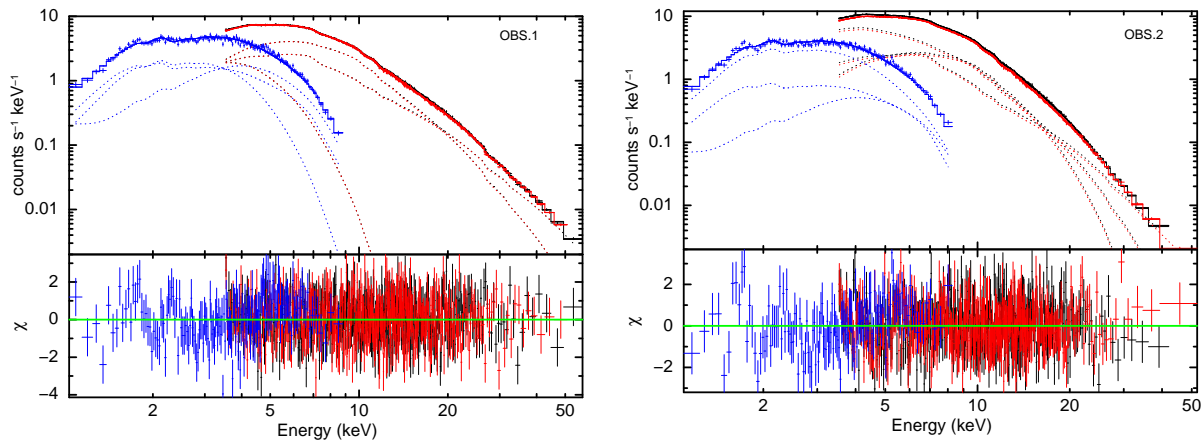

\centering
\includegraphics[scale=0.32,angle=-90]{fig6a.ps}
\includegraphics[scale=0.32,angle=-90]{fig6b.ps}
\caption{Unfolded energy spectra of 4U~1728--34 during both the observations are shown in both panels where \nustar{} and \swift{} spectra are jointly fitted using the best-fit model {\tt TBabs$\times$(bbody+CompTT+relxill)}. Bottom panel shows residuals of the fitted spectra for OBS 1 and OBS 2, respectively. }
  \label{reflection_ratio}
\end{figure*}

The model {\tt TBabs$\times$(bbody+CompTT+relxill)} resulted in $\chi^{2}/dof=1098/940$ and $937/836$ for the OBS 1 and OBS 2, respectively. Fitted spectra along with model components and residuals are shown in the left and right panel of Fig.~\ref{reflection_ratio}. We fixed the outer radius $R_{out}=1000R_{G}$ as the sensitivity of the reflection fit goes down to a larger value of the outer disc radius. The source has a known spin frequency of $\sim 363$ Hz \citep{1996ApJ...469L...9S}. The dimensionless spin parameter $\it a$ can be calculated from the spin frequency using the relation $a\simeq0.47/P[ms]$ \citep{2000ApJ...531..447B}. Since $\nu=363$ Hz for the source 4U~1728--34, we fixed $\it a$ at 0.17 while fitting. As the inner radius for several other NS LMXBs were found in the range $5-20R_{G}$ (e.g. \citealt{2010ApJ...720..205C, 2011A&A...530A..99E}), we set the breaking radius larger than this range at $R_{br}=25R_{G}$. 

\begin{figure*}
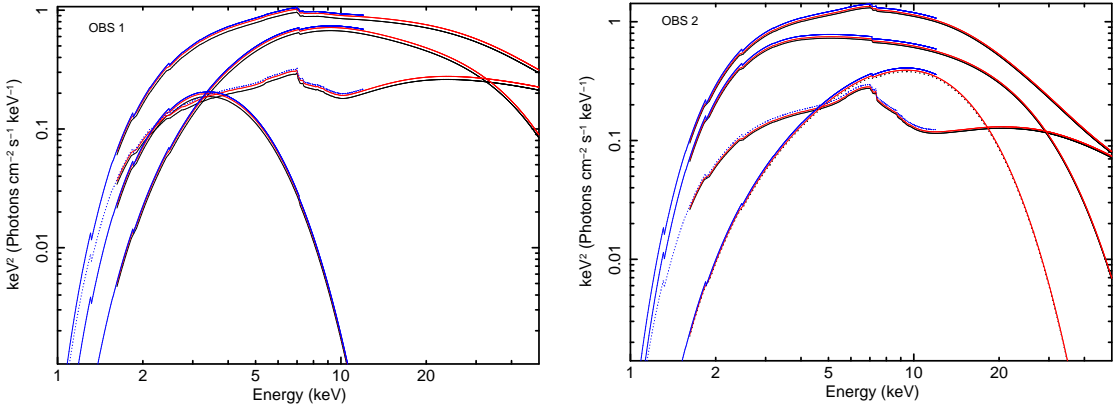

\centering
\includegraphics[scale=0.30,angle=-90]{fig7a.ps}
\includegraphics[scale=0.30,angle=-90]{fig7b.ps}
\caption{Best fit model spectra used to fit joint \swift{} and \nustar{} observations of 4U~1728--34 are shown for OBS 1 (left panel) and OBS 2 (right panel). Relative contribution of individual model components $-$ {\tt bbody}, {\tt compTT} and {\tt relxill} are also shown.}
  \label{modelspec}
\end{figure*}

\begin{figure*}
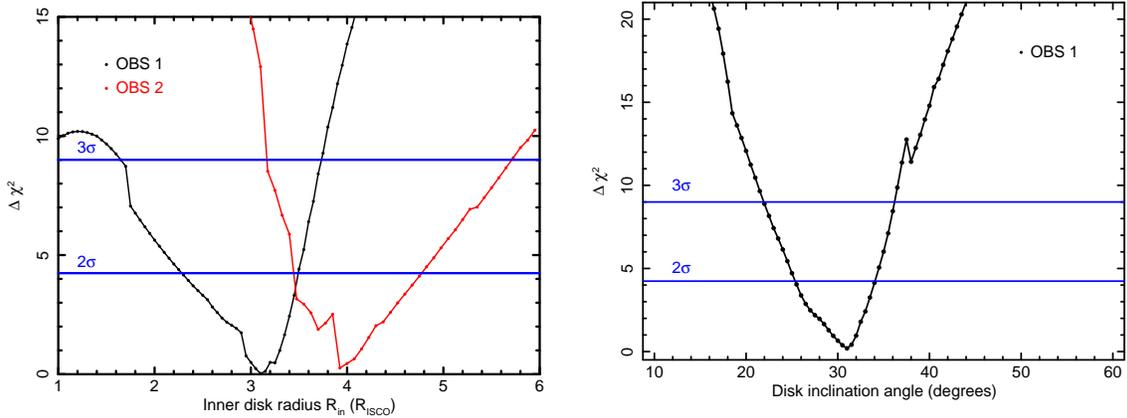

\centering
\includegraphics[scale=0.30,angle=-90]{fig8a.ps}
\includegraphics[scale=0.30,angle=-90]{fig8b1.ps}
\caption{Employing Delta statistics, the variation in $\Delta\chi^2$ = $\chi^2$ - $\chi^2_{min}$ for different values of the best-fit model parameter is shown. Left panel shows $\Delta\chi^2$ variations as a function of inner disc radius (in the unit of $R_{isco}$; obtained from {\tt relxill} model) as observed from OBS 1 (black) and OBS 2 (red) and right panel shows the same as a function of inner disc inclination angle for OBS 1 (obtained from {\tt relxill} model). Horizontal lines in both panels indicate 2$\sigma$ and 3$\sigma$ significance level.}
  \label{delchi}
\end{figure*}

The obtained inner disc radius lies at $R_{in}\simeq(2.8-4.3)\times R_{ISCO}$. We found a moderately high ionization parameter of log$\xi\simeq3.3-3.7$, which may be expected for a broad Fe line. From both the observations, our spectral fit resulted in relatively flat power-law with $\Gamma\leq2.0$. From the OBS 1, we inferred moderately low disc inclination angle of $i\simeq32^{0}$ but the same was fixed to the value $32^{0}$ for OBS 2. Iron abundance with respect to the solar value was found to be in the range $A_{Fe}\simeq2-5$ and consistent from both observations within error-bars. Model parameters of the best fit spectra for OBS 1 and OBS 2 are summarized in Table~\ref{fit3} along with the 2$\sigma$ errors. Best fit model spectra for both the observations are shown in the left and right panels of Fig.~\ref{modelspec} along with individual model components. In order to constrain inner disc radius as well as disc inclination angle from our best-fit model, we computed $\Delta\chi^2$ for each of the parameters using {\tt steppar} in {\tt XSPEC}. The resultant $\Delta\chi^2$ while varying inner disc radius as the free parameter between $1\: R_{ISCO}$ and $6\: R_{ISCO}$ for the OBS 1 and OBS 2 are shown in the left panel of Fig.~\ref{delchi}. The right panel shows the resultant $\Delta\chi^2$ while varying the disc inclination angle in OBS 1 between 10$\degree$ and 60$\degree$. In both panels, 2$\sigma$ and 3$\sigma$ significance levels are shown by horizontal lines. Within 3$\sigma$ bounds, the disc inclination angle from OBS 1 is found to be $31^{+6}_{-9}$ degree while the inner disc radius are found to be $3.1^{+0.5}_{-1.2}$ and $3.9^{+1.6}_{-0.7}\: R_{ISCO}$ for the OBS 1 and OBS 2, respectively.  \\

\subsubsection{Complex absorption}
From the spectral analysis based on \chandra{} and \rxte{} or \sax{} data \citet{2005astro.ph.11072D} showed that the Iron line complex at $\sim6.5\kev$ may also be well described with two absorption edges. Following \citet{2005astro.ph.11072D}, \citet{2011A&A...530A..99E} also employed two absorption edges to fit the excess seen $\sim6.5\kev$ in the \xmm{} spectra of 4U~1728--34.
 In this work, We observed a significant improvement of the fit when we employed two smeared edge model ({\tt smedge} in {\tt XSPEC}) to the continuum model {\tt TBabs$\times$(bbody+CompTT)}. Our spectral fitting required two smeared edges to fit the broad residuals seen in the Fe K band. The model {\tt TBabs$\times$smedge$\times$smedge(bbody+CompTT)} resulted in $\chi^{2}/dof=1129/943$ and $1004/839$ for the OBS 1 and OBS 2, respectively with the edge energies $E_{edge}=6.81\pm0.06\kev$ ($\tau\simeq1.33$) and $8.97^{+0.30}_{-0.25}\kev$ ($\tau\simeq0.20$) for the OBS 1 and $E_{edge}=7.02\pm0.06\kev$($\tau\simeq1.39$) and $9.0^{+0.18}_{-0.20}\kev$ ($\tau\simeq0.64$) for the OBS 2, respectively. The {\tt smedge} widths were not well constrained for OBS 2 and we have frozen those to the value $10 \kev$.  However, it may be noted that modelling with the reflection model {\tt relxill} provided us statistically better fit compared to the two {\tt smedge} models. 

\begin{table*}
\centering
\caption{Best-fit parameter values of the individual burst spectra as well as the combined burst spectrum in the energy band $3-20\kev$} 
  \begin{tabular}{|p{2.5cm}|p{1.5cm}|p{1.5cm}|p{1.5cm}|p{1.5cm}|p{1.5cm}|}
    \hline
    \hline
    Parameter & Burst 1 & Burst 2 & Burst 3 & Burst 4 & Combined  \\ 
    \hline
    $N_{H}(\times10^{22}$ cm$^{-2}$) & 2.5 (f) & 2.5 (f) & 2.5 (f) & 2.5 (f) & 2.5 (f)\\
    $kT_{bb}$ (\kev)......&$2.27\pm0.07$ & $2.20\pm0.08$ & $2.11\pm0.07$ & $2.11\pm0.10$ &$2.22\pm0.05$ \\
    $R_{NS}$ (km) ............ & $8.09\pm1.65$&$8.67\pm1.76$ & $10.41\pm2.11$&$9.54\pm1.79$ & $9.87\pm1.86$ \\
    $\chi^{2}/dof$............&$101/103$ &$128/106$ &$160/111$ &$115/99$ & $217/181$ \\   
    \hline 
  \end{tabular}\label{burst_spectra} \\
\begin {tablenotes}
\small
 \item $N_H$ is the neutral Hydrogen absorption column density, $kT_{bb}$ is the blackbody temperature, $R_{NS}$ is the radius of blackbody emission and $\chi^{2}/dof$ is the reduced $\chi^2$ of the best fit. The color correction factor and the distance to the source are assumed to be 1.7 and 5 kpc, respectively.
\end {tablenotes}
\end{table*}

\begin{table*}
   \caption{ Best-fit spectral parameters of the combined \swift{} and \nustar{} spectra for both the observations using model: {\tt TBabs$\times$(bbody+compTT+relxill)}. Quoted errors are at $90\% $ confidence level.} 
\begin{tabular}{|p{2.0cm}|p{3.5cm}|p{2.0cm}|p{2.0cm}|}
    \hline
    \hline
    Component     & Parameter & \multicolumn{2}{c}{\swift{}+\nustar{}} \\
     \cline{3-4}
                   &          & OBS 1 & OBS 2 \\ 
   \hline
    {\scshape tbabs}    & $N_{H}$($\times 10^{22} cm^{-2}$) &$3.91\pm0.12$ & $4.05_{-0.09}^{+0.22}$    \\
    {\scshape blackbody} & $kT_{bb}$($\kev$) & $0.71\pm0.05$ & $2.40\pm0.02$  \\
    & $N_{BB}$~$^a$($\times 10^{-2}$)   & $0.63_{-0.14}^{+0.04}$  &  $1.24_{-0.17}^{+0.07}$    \\

    {\scshape comptt} & $kT_{seed}(\kev)$ & $1.46\pm0.10$ & $0.53\pm0.03$    \\
    & $kT_{e}$ (keV) & $7.1_{-1.5}^{+7.0}$ &  $4.6_{-0.6}^{+0.3}$ \\
    & Optical depth($\tau$) & $2.39_{-0.87}^{+0.77}$  &  $3.32\pm0.40$ \\
    
    & $n_{comptt}$~$^b$  & $0.02\pm0.01$ & $0.15_{-0.01}^{+0.04}$    \\ 
   {\scshape relxill} & Index($\beta 1$) & $>3$ & $>3$   \\
   & Index($\beta 2$) & $3(f)$ & $3(f)$    \\
   & $i(^0)$ & $31_{-2}^{+3}$ & $32(f)$    \\
   & $R_{in}$($\times R_{ISCO}$) & $3.1_{-0.3}^{+0.5}$ & $3.9_{-0.6}^{+0.4}$ \\
   & $\Gamma$ & $\leq 1.96$ & $\leq 1.45$ \\
   & log$\xi$ & $3.3\pm0.2$ & $3.7\pm0.1$ \\
   & $A_{Fe}$($\times$ Solar) &$2.1_{-1.0}^{+3.5}$  & $4.9\pm1.0$ \\
   & $R_{refl}$ & $1.1_{-0.7}^{+1.4} $ & $ 1.0_{-0.1}^{+0.4}  $ \\
   & $N_{refl}$~$^c$ & $0.11\pm0.06$ & $0.05\pm0.01$ \\
   & $F_{total}$ ($10^{-9}$ ergs/s/cm$^2$) & 3.9 $\pm$ 0.2 & 5.1 $\pm$ 0.3 \\	
   & $F_{bbody}$ ($10^{-9}$ ergs/s/cm$^2$) & 0.5 $\pm$ 0.1 & 1.2 $\pm$ 0.1 \\
   & $F_{comptt}$ ($10^{-9}$ ergs/s/cm$^2$) & 1.8 $\pm$ 0.1 & 2.6 $\pm$ 0.2 \\
   & $F_{relxill}$ ($10^{-9}$ ergs/s/cm$^2$) & 1.6 $\pm$ 0.1 & 1.3 $\pm$ 0.1 \\
   
\hline 
    & $\chi^{2}/dof$ &$1098/940$  & $937/836$   \\
    \hline
  \end{tabular}\label{fit3} \\
\begin{tablenotes} 
\small
\item The outer radius of the {\tt relxill} spectral component was fixed to $1000R_{G}$ and the spin parameter was set to $a=0.17$. Breaking radius was set to $R_{br}=25 R_{G}$. ~$^{a,b,c}$ denotes the normalization component of the {\tt bbody}, {\tt compTT} and {\tt relxill} model, respectively.  
\end{tablenotes}
\end{table*}

\section{Discussion}
In this work, we used two \swift{}/XRT and \nustar{} simultaneous observations of the Atoll-type NS LMXB 4U~1728--34 with a gap of one day and performed both the timing and spectral analysis. Using power density spectral study, hardness ratio and colour-colour diagram, we showed that the OBS 1 belongs to the island state while the OBS 2 belongs to the lower banana state. We observed a LF QPO at $\sim$4.53 Hz which is a signature of the island state. Later broadband spectral analysis of both observations in the energy range $1-79 \kev$ also suggests that the OBS 2 is relatively softer than the OBS 1, which is consistent with dual spectral states of both observations. 

\subsection{Burst emission}
We observed four type-I X-ray bursts from the OBS 1. Interestingly after an initial rise, they show different morphologies. Fig.~\ref{burst_lc} shows that bursts with high peak count rate show sharp fall in intensity as soon as they reach the peak count rate. However, bursts with low peak count rate spent $\sim2-3$ sec at the peak and then they follow exponential fall. Burst profiles with low peak counts are typical evidence of photospheric radius expansion where the radius expansion drives the photosphere to large radii \citep{2008ApJS..179..360G, 2010ApJ...724..417G}. Therefore, it is evident that true critical luminosity was not achieved in this source during photospheric radius expansions and brightest bursts from our observations do not show photospheric radius expansion. However, the reason of such alternating occurrence of photospheric radius expansion X-ray bursts and bright X-ray bursts profiles is not well understood. Therefore, considering individual bursts as well as combined burst, we performed burst spectral analysis in the energy range $3-79\kev$. The combined burst spectral analysis provided the effective blackbody temperature of $\sim2.2\kev$. The measured color temperature and the effective temperature are related by, $T_{col}=fT_{eff}$, where $f$ is the spectral hardening factor which may range over $1.2-1.7$ \citep{1984ApJ...287L..27L}. If we assume the distance of the source $\sim 5$ kpc and spectral hardening factor $1.7$, then the effective radius of the emitting region becomes $\simeq9.87$ km (using the relation, $R_{eff}=f^{2}R_{in}$), which is comparable to the radius of the neutron star \citep{2000ApJ...542.1034D, 2003ApJ...593L..35S}.

\subsection{Persistent emission}
In the persistent spectra, we described the continuum emission from the LMXB 4U~1728--34 with a blackbody emission from the NS surface together with a Comptonized emission from an extended corona or the boundary layer. Joint \swift{} and \nustar{} spectra detected a broad emission feature centred at energy $\sim 6.5\kev$, compatible to a fluorescent Fe K$\alpha$ line. The observed Fe K$\alpha$ line has a broadness of $\sigma/E\simeq0.1$.
The broadness of the line suggests the reflection of hard X-ray photons in the accretion disc where the strong velocity field broadens discrete features and relativistic effects distort their shapes. In the observation analysed here, the relativistic reflection model {\tt relxill} successfully modelled the Iron line complex that includes both line as well as reflection hump features. Our spectral fit estimated an inner accretion disc radius $R_{in}\approx (2.8-4.3)\:R_{ISCO}$. For a spinning neutron star with $a=0.17$, $R_{ISCO}$ can be approximated as $R_{ISCO}\simeq (6\:GM/c^2)(1-0.54a)$ \citep{2000arxt.confE...1V}. So, the inner accretion disc radius would correspond to $R_{in}=(15-23)\:GM/c^2=32-48$ km for a mass of $1.4\:M_{\odot}$ (as $R_{ISCO}\simeq 5.4\:GM/c^2$ here). Our estimate of the inner disc radius is compatible with the value of $25-100$ km inferred by \citet{2011A&A...530A..99E} after modelling a \xmm{} spectrum of this source. The obtained value of the inner disc radius suggests that the disc would be truncated moderately far away from the neutron star surface. During the \nustar{} observation, the source was observed in the low/hard state for which the disc should be truncated far from the compact object. Moreover, the Keplerian frequency associated with this truncation radius was found to be in the range $210-386$ Hz, which is consistent with the observed spin frequency ($\sim 363$ Hz) by \citet{1996ApJ...469L...9S}. It may be noted that the orbital frequency is calculated by using the following relation \citep{2000arxt.confE...1V}
$$\nu_{orb}\approx 1200(r_{orb}/15\:\text{km})^{-3/2} m_{1.4}^{1/2}\,\text{Hz} $$
where $m_{1.4}$ is the mass of the NS in units of $1.4M_{\odot}$. The system inclination with respect to the line of sight was estimated to be $31^{+6}_{-9}$ degree with 3$\sigma$ significance. The measured inclination is consistent with this system as no X-ray dips were observed in its \nustar{} light curve. While modelling reflection features, we observed that while going from hard island state to relatively soft lower banana state, the reflected flux decreases as well as the ionization parameter increases from log$\xi=$ 3.3 to 3.8 although the width of the Iron line complex does not change. High ionization parameter is also consistent with \citet{2011A&A...530A..99E}. A closer inspection of Fig.~\ref{residuals} reveals that the Iron line complex from both observations has a broad blue wing in addition to the broad red wing of the line. This makes the line width unusually broader and symmetric than usual asymmetric line profile. Such large width of the Iron line can be explained by processes like Compton broadening in a hard corona \citep{2009MNRAS.399L...1R}. To produce the observed line width of $\sim0.7\kev$, the electrons in the Comptonizing cloud should have a temperature of $\simeq15\kev$ \citep{1979rpa..book.....R}. Our best-fit model estimated an electron temperature (3kT$_e$ = $13-21 \kev$ from Table~\ref{fit3}) is very similar to that of the proposed value. Optically thin boundary layers are often considered to be the origin of such Comptonization. Therefore, if such Comptonized photons are the source of incident photons for reflection, a symmetric broadening is expected. Such a broad, symmetric Iron line profile have also been observed from the \nustar{} spectrum of the source Aql X-1 \citep{2016ApJ...819L..29K}. However, further investigation of the origin of such broadening is out of the scope of the present work.   \\

From both observations, our best-fit spectral model provided an unabsorbed bolometric flux in the $0.001-100\kev$ band is $\simeq3.87-5.05\times10^{-9}$ ergs cm$^{-2}$ s$^{-1}$, which corresponds to the X-ray luminosity of $L_{X}\simeq1.1-1.6\times10^{37}$ ergs s$^{-1}$ for an assumed distance of $5$ kpc \citep{2000ApJ...542.1034D}. This luminosity is $6-9\%$ of the Eddington luminosity, typical for atoll sources ($L_{Edd}\simeq 1.76\times10^{38}$ ergs s$^{-1}$ for a $1.4M_{\odot}$ neutron star). Therefore, accretion rate is moderate in both island and lower banana branch observations. Moderate accretion rate during these two branches was also observed from this source by \citet{2007ApJ...654..494T}. It is generally believed that during the high/soft states, when the accretion rate is high ($\sim10\%$ of the $L_{Edd}$), the disc extends close to the compact object, whereas in the quiescence the inner disc radius truncates far from the compact object ($\sim100 GM/c^{2}$). From the inferred accretion rate as well as the inner disc radius, it is expected that our observation lies in the boundary of hard to soft X-ray spectral state transition or more accurately transiting from island state to lower banana state.
We found that between the two observations analyzed here, the temperature of the NS (which is equivalent to the blackbody temperature) increases, whereas the electron temperature of the Comptonizing corona decreases as the source moves from the harder island state (OBS 1) to the softer lower banana state (OBS 2). \\

The detection of the phenomenon like thermonuclear X-ray burst and quasi-periodic oscillations suggest that the magnetic field of the NS in LMXBs is not very strong (may lie in the range $10^8-10^{10}$ G ;\citealt{2013MNRAS.429.3411P}). X-ray pulsations from the accreting NS are observed when the NS magnetic field is strong enough to channel down some fraction of the accreted matter from the NS surroundings to the magnetic poles. So far X-ray pulsations from the source 4U~1728--34 has not been significantly detected, but frequent type-I X-ray bursts have been observed from the NS surface. Additionally, from the OBS 1 and OBS 2 we calculated the inner disc radius to be 3.1$^{+0.5}_{-1.2}$ and 3.9$^{+1.6}_{-0.7}$ $R_{ISCO}$ respectively with 3$\sigma$ significances. This is fairly large truncation radius of the order of $\sim$30-50 km. It has been observed that in some LMXBs the disc truncation occurs at moderate radii due to the pressure exerted by the magnetic field of the NS \citep{2014ApJ...796L...9D} or due to evaporation of the inner disc at low accretion rate \citep{2013MNRAS.429.3411P}. We explored both possibilities follows.
 To estimate the magnetic field, we used the following relation given by \citet{1975A&A....39..185I}
$$ R_{in}=4\times10^{8}B_{11}^{4/7}\dot m_{15}^{-2/7} M^{-1/7} \text{cm} $$
where $B_{11}$ is the magnetic field in units of $10^{11}$ G, $\dot m_{15}$ is the mass accretion rate in units of $10^{15}$ gm s$^{-1}$ and $M$ is the mass of the NS in $M_{\odot}$ units. The average mass accretion rate for this source is $2.6\pm1.6\times10^{-9} M_{\odot}$ yr$^{-1}$ \citep{2013ApJ...768..184H}. Using the measured inner disc radius from the reflection model and assuming the mass of the NS to $1.4M_{\odot}$, we estimated a magnetic field strength of $B\approx(3.3-6.5)\times10^{8}$ G.
As the calculated magnetic field strength of the NS in this binary is not very high ($\sim10^{8}$ G) and the accretion rate during the persistent emission onto the compact object is moderate ($\sim5-8\%$ of the Eddington limit). Therefore, the second possibility that the evaporation occurs at the inner disc at moderate accretion rate may be the favourable one with the observed moderate disc truncation scenario in our case. \\ 

We found that the inner disc is truncated moderately far from the NS surface. Even if the total bolometric flux increases from island state to lower banana state, the inner disc remains at similar truncation radius within 3$\sigma$ limits. At the same time \nustar{} observation showed the evidence of type-I X-ray burst which requires dumping of accreting material onto neutron star surface. It essentially indicates that the accreted material is still reaching to the surface of the NS in spite of moderate disc truncation. This behaviour can be explained with the model given by \citet{1991ApJ...372L..87K}. According to this model, the accreted material can free fall crossing the `gap' between the disc and the NS surface and then strikes the NS surface, creating a hot accretion belt with a temperature inversion.    
 
\section{Acknowledgements}
This research has made use of data and/or software provided by the High Energy Astrophysics Science Archive Research Centre (HEASARC). Aditya S. Mondal would like to thank Inter-University Centre for Astronomy and Astrophysics (IUCAA) for hosting him during subsequent visits.

\def\apj{ApJ}
\def\apjl{ApJl}
\def\pasp{PASP} \def\mnras{MNRAS} \def\aap{A\&A} \def\physerp{PhR} \def\apjs{ApJS} \def\pasa{PASA}
\def\pasj{PASJ} \def\nat{Nature} \def\memsai{MmSAI} \def\araa{ARAA} \def\iaucirc{IAUC}
\bibliographystyle{mn2e}
\bibliography{aditya}

\end{document}